\documentclass[pra,%
 reprint,
 amsmath,amssymb,
 aps,
]{revtex4-2}

\usepackage{graphicx}% Include figure files
\usepackage{dcolumn}% Align table columns on decimal point
\usepackage{bm}% bold math
\usepackage{mathrsfs}
\usepackage{amsfonts}
\usepackage{amssymb}
\usepackage{amsmath}
\usepackage{epsfig}
\usepackage{bbold}
\usepackage{subfigure}
\begin{document}

\preprint{APS/123-QED}

\title{Classical and Quantum Resources in Perfect Teleportation}% Force line breaks with \\

\author{Dian Zhu}
 \affiliation{Theoretical Physics Division, Chern Institute of Mathematics, Nankai University, Tianjin 300071, China}%Lines break automatically or can be forced with \\
 
\author{Fu-Lin Zhang}
 \email{flzhang@tju.edu.cn}
 \affiliation{Department of Physics, School of Science, Tianjin University, Tianjin 300072, China}
 
\author{Jing-Ling Chen}
 \email{chenjl@nankai.edu.cn}
 \affiliation{Theoretical Physics Division, Chern Institute of Mathematics, Nankai University, Tianjin 300071, China}

\date{\today}% It is always \today, today,

\begin{abstract}
%我们提出了一个新的隐形传态方案，该方案是用部分纠缠的two-qutrit信道来完美传递一个qubit。
%%%%%%%%%%%Novel作为“新颖的”意思，强调前所未有的创新性或独特性，用在这里不好。用new即可。
We propose a teleportation protocol that enables perfect transmission of a qubit using a partially entangled two-qutrit quantum channel.
%利用我们的方案，我们分析了隐形传态的三个关键因素——量子信道，由发送方执行的测量和发送给接收方的经典信息——之间的关系。
Within our scheme, we analyze the relationship among the three key ingredients of teleportation: 
(i) the quantum channel,
(ii) the sender's (Alice's) measurement operations, and
(iii) the classical information transmitted to the receiver (Bob).
Compared to Gour's protocol \cite{PRA2004}, our scheme requires less entanglement of Alice's measurement and fewer classical bits sent to Bob.
%我们的结果还显示了这两种资源之间的权衡，并推导了它们之和的上界，量化了它们在隐形传态过程中的相互作用/影响。
Our results also show a trade-off between these two resources and derive a lower bound for their sum, quantifying their interplay in the teleportation process.
\begin{description}
\item[Keywords]
teleportation; entanglement; classical information;
\end{description}
\end{abstract}

\maketitle

%\tableofcontents

\section{\label{sec:level1}Introduction}

%第一段介绍teleportation的相关背景，发展方向等内容。

%量子隐形传态是一个这样的过程，利用一个量子信道，在经典通信的帮助下将未知的量子比特从发送方传递到接收方。
Quantum teleportation is a protocol that enables the transmission of an unknown quantum state from a sender (Alice) to a receiver (Bob), utilizing a quantum channel and assisted by classical communication.
%这个过程最初由Bennett等人提出，其完美实现依赖于在Alice和Bob之间共享最大纠缠的两量子比特贝尔态作为量子信道。
This process, first proposed by Bennett et al. \cite{bennett1993teleporting}, can be perfectly implemented by sharing a maximally entangled two-qubit Bell state between Alice and Bob.
%简单来说，Alice通过对待传递的量子态和预共享的Bell态的子系统进行联合测量，并将相应的测量结果通过经典通讯传递给Bob，Bob最后将手中的量子态利用某些门操作以100percent的概率来恢复目标状态。
Alice performs a joint measurement on the quantum state to be transmitted and a subsystem of the pre-shared Bell state.
She then sends the corresponding measurement results to Bob via classical communication.
Finally, Bob applies specific unitary operations to his state, recovering the target state with $100\%$ success probability.
%除了这个原始方案之外，研究人员还将隐形传态推广到概率实现（例如，使用部分纠缠的纯态）和高维量子信道，大大拓宽了其理论范围。
%Beyond this original scheme, researchers have generalized teleportation to probabilistic implementations (e.g., using partially entangled pure states) and high-dimensional quantum channels, significantly broadening its theoretical scope \cite{li2000probabilistic,banaszek2000optimal,roa2003optimal,verstraete2003optimal,roa2015probabilistic,luo2019quantum,huang2020quantum,hu2020experimental,PRA2004,chen2022perfect,chen2022probabilistic}.

%第二段本文研究问题的来源，即为什么要研究这个问题，该问题在teleportation中的重要性。

%隐形传态的一大特点是其协议的不唯一性。利用部分纠缠的纯态来概率实现隐形传态，用高维量子系统来实现完美隐形传态，以及涉及第三方作为控制者的受控隐形传态等，都是来自最初版本隐形传态的推广，并在量子通信的多种场景中扮演关键角色。
A key feature of quantum teleportation lies in the non-uniqueness of its protocols. Significant extensions beyond the original scheme include: probabilistic teleportation with partially entangled pure states~\cite{li2000probabilistic,banaszek2000optimal,roa2003optimal,verstraete2003optimal,roa2015probabilistic},
teleportation in high dimensions~\cite{luo2019quantum,huang2020quantum,hu2020experimental,chen2022perfect,chen2022probabilistic},
and controlled teleportation involving third-party supervisors~\cite{karlsson1998quantum,li2014control}.
These protocol variants now play a pivotal role in various quantum communication scenarios~\cite{sangouard2011quantum,biham1996quantum,bose1998multiparticle,townsend1997quantum,quantnet2004}.
%尽管协议存在多样性，但这些协议都存在一个共同点，即需要付出量子纠缠资源和经典通信资源作为代价。
Despite the diversity of protocols, they all share a common requirement: the consumption of quantum entanglement and classical communication resources.
%前者体现为传输信道的纠缠度与发送方联合测量所需的纠缠资源，后者则取决于重构量子态所需传输的经典信息量。
%The performance of quantum teleportation relies on utilizing two essential resources: quantum entanglement and classical communication.
The former manifests in the entanglement of the quantum channel and the entanglement resources of the sender's joint measurement, while the latter depends on the amount of classical information that must be transmitted to reconstruct the quantum state.
%这三个关键因素之间存在什么样的关系一直是隐形传态理论研究的一个开放问题。特别地，当量子信道固定时，测量所需的资源与经典通信所需的资源存在什么样的关系？这一关系能否被量化？
The interplay among these three key factors remains an open problem in quantum teleportation theory. Specifically, for a fixed quantum channel, what is the fundamental relationship between the measurement resource requirement and the classical communication cost? Can this relationship be quantitatively characterized?

A qubit is the basic unit of quantum information. In its simplest teleportation scheme, the channel is a maximally entangled two-qubit state, Alice performs a maximally entangled measurement, and two classical bits are transmitted. These three ingredients are fixed, leaving no freedom to expose their interrelations.
By extending to two-qutrit channels, such freedom emerges, and our results show that when channel entanglement exceeds one, measurement entanglement is typically reduced, meaning the difficulty of measurement can be traded for that of state preparation.
%我们在利用任意部分纠缠的两个量子纯态忠实传输量子比特的特殊情况下讨论了这些问题。
We address these questions for the special case of faithfully transmitting a qubit using arbitrary partially entangled two-qutrit pure states.
%由于通过量子比特信道完美传输量子比特的必要且充分条件是信道本身必须是最大纠缠态——这种理想资源在实验上既难以制备，又难以长时间保持——因此使用三能级（qutrit）信道具有显著优势：即使是非最大纠缠的两三能级态，也能够实现量子比特的忠实传输。在此背景下，我们将 Alice 的测量基解释为计算基的一种幺正变换，并将量子隐形传态协议从两量子比特信道扩展到两三能级信道。
%%%%%%%%%%%%%%%%%%%%%%%Since the necessary and sufficient condition for perfectly transmitting a qubit through a qubit channel is that the channel itself be maximally entangled, which is an ideal resource that is experimentally difficult both to generate and to preserve for long durations, the use of qutrit channels provides a notable advantage: even non-maximally entangled two-qutrit states can still enable faithful qubit transmission.
%%%%%%%%%%%%%%%%%%%%%%%Building on this motivation, we interpret Alice’s measurement basis as a unitary transformation of the computational basis and extend the teleportation protocol from two-qubit to two-qutrit channels.
%我们的方案采用五个自由参数来描述测量基础，与现有协议相比，提供了更广泛的适用性：
Our scheme employs five free parameters to characterize the measurement basis, offering broader applicability compared to existing protocols: 
%Gour的协议只需要两个参数，但不管信道纠缠度如何，都会对log2施加固定的经典代价。
Gour's protocol~\cite{PRA2004} requires only two parameters but imposes a fixed classical cost $\log_2 d_a$ (where $d_a$ represents the total dimension of Alice's system), regardless of channel entanglement.
%Chen等人的协议进一步将参数计数减少到1，尽管这是以限制为具有简并Schmidt系数的部分纠缠态为代价的。
Chen et al.'s protocol~\cite{chen2022perfect} further reduces the parameter count to one, though this comes at the expense of being restricted to partially entangled states with degenerate Schmidt coefficients.
In contrast, our protocol combines these two characteristics: (i) entanglement-dependent classical communication cost for resource optimization, (ii) compatibility with adaptability to arbitrary partially entangled two-qutrit channels.
Based on Ref.~\cite{chen2022perfect}'s definition of the entanglement of Alice's joint measurement basis, we analyze the relationship among the above three key ingredients.
%研究结果表明，与现有方案相比，本协议在所有满足完美传态条件的部分纠缠两qutrit纯态下，均在经典通信成本与量子纠缠消耗两方面实现了更优的资源效率。更重要的是，我们推导出固定信道条件下这两类资源总消耗的下界，揭示了经典与量子资源之间受约束的权衡关系。
The results show that our protocol achieves greater resource efficiency than prior approaches~\cite{PRA2004,chen2022perfect} in both classical communication costs and quantum entanglement consumption for any perfect-teleportation-enabled two-qutrit partially entangled pure state.
Moreover, we derive the lower bound on the sum of these two resource costs under fixed channel conditions, revealing a constrained trade-off relationship between classical and quantum resources.

\section{Two-qubit channel}

%为了提出一种新的用于量子比特传输的双量子信道协议，我们重构了原先为双量子比特信道设计的标准隐形传态方案。
%首先，我们简要回顾了利用两个量子比特贝尔态作为量子通道的传统方法。
To propose a  two-qutrit channel protocol for qubit transmission, we reconstruct the standard teleportation scheme originally designed for two-qubit channels.
First, we briefly revisit the conventional approach that utilizes two-qubit Bell states as the quantum channel.

\begin{figure}[htbp]
  \centering
  \includegraphics[width=0.4\textwidth]{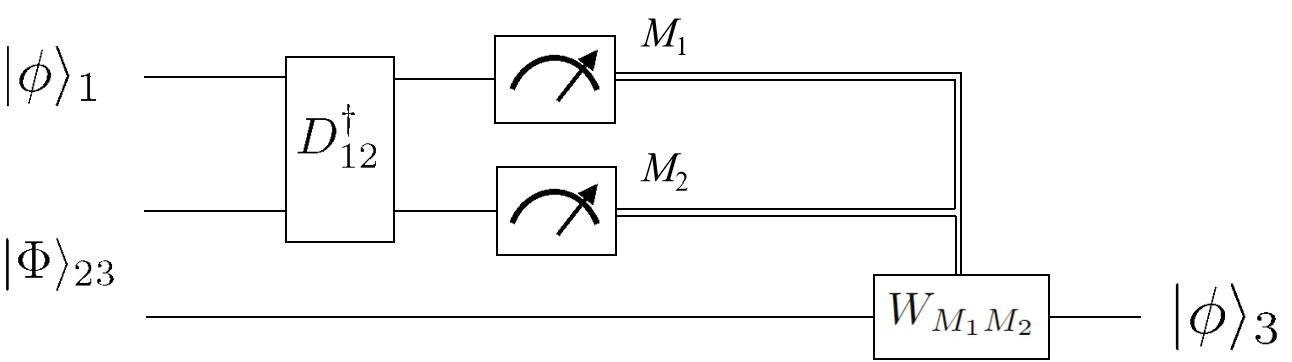}
  \caption{
  The quantum circuit diagram shows that qubit teleportation is realized with Alice's (top two wires) and Bob's (bottom wire) subsystems. The single lines denote qubits, and the double lines denote classical bits. %Alice performs a unitary operation $D_{12}^\dagger$ and follows the measurements on qubits 1 and 2 independently. After obtaining Alice's measurement outcomes $M_1$ and $M_2$, Bob can recover the initial state by performing appropriate unitary operations $W_{M_1 M_2}$ to qubit 3 in his hands.
  }  \label{Fig1}
\end{figure}

%根据标准协议，发送方和接收方预共享一对Bell态。假定Alice需要传递一个未知的单比特量子态。
Following the standard protocol, Alice and Bob pre-share an EPR pair.
To transmit an unknown qubit
\begin{equation}  \label{deliverstate}
    |\phi\rangle_1 = \alpha |0\rangle_1 + \beta|1\rangle_1
\end{equation}
($|\alpha|^2 + |\beta|^2 = 1$), Alice applies a controlled-Not gate (qubit 1 as control, qubit 2 as target) and a Hadamard on qubit 1,
%Here, $H$ denotes the Hadamard gate acting on qubit 1, and $U_{CNOT}$ represents the controlled-NOT gate with qubit 1 as the control and qubit 2 as the target.
then measures qubits 1 and 2 in the computational basis, yielding outcome $m \in \{00, 01, 10, 11\}$.
Based on $m$, Bob applies the corresponding Pauli gate ($\mathbb{1}$, $X$, $Z$ or $Y$) to qubit 3, perfectly recovering $|\phi\rangle_1$.
%图一展示了标准隐形传态的流程图，其中$|\Phi\rangle_{23}$是预共享的EPR对，$D_{12}^\dagger$是Alice执行的幺正操作，$W_{M_1 M_2}$是Bob根据得到的经典信息所执行的恢复操作。
Figure~\ref{Fig1} illustrates this standard teleportation protocol, where $|\Phi\rangle_{23}$ is the pre-shared EPR pair, $D_{12}^\dagger$ Alice's unitary, and $W_{M_1 M_2}$ represents Bob's recovery operation using the classical outcomes.

Notice that the pre-shared EPR pair can be replaced by a two-qubit partially entangled pure state
\begin{equation} %\label{}
  |\Phi\rangle_{23} = a_0 |00\rangle_{23} + a_1 |11\rangle_{23}
\end{equation}
with $\sum_{j=0}^1|a_j|^2 = 1$, where perfect teleportation is attainable if and only if $|\Phi\rangle_{23}$ is maximally entangled~\cite{li2000probabilistic}.
%在我们后续的分析中，将采用上述形式及其高维推广作为量子信道的状态。
Our subsequent analysis will utilize such partially entangled states (and their high-dimensional extensions) as the quantum channel.
Without loss of generality, one can assume that the Schmidt coefficients $a_{j=0,1}$ are real numbers and $0 \leq a_0 \leq a_1$.
Meanwhile, the unitary operation $D_{12}^\dagger$ and the following measurement of the computational basis can be regarded as a measurement of the Bell basis. This implies that constructing Alice's joint measurement basis reduces to designing the unitary operator $D_{12}$.
%由于Bob要执行的恢复操作依赖于算符$D_{12}$的元素。因此，隐形传态协议的核心步骤是对幺正算符$D_{12}$的设计。
Since Bob's recovery operation depends on the elements of operator $D_{12}$, the essential step of the teleportation protocol lies in the design of this unitary operator.
%Hence, the essential step of the teleportation protocol is the unitary transformation of Alice's system.

%在我们的方案下，Alice执行的联合测量对应的幺正算符形如
In our scheme, the unitary operator corresponding to Alice's joint measurement is of the form
%The unitary in our scheme is of the form
\begin{equation}
    D_{12} \!\!=\!\! \left(\!\!
    \begin{array}{cccc}
        u_{11} & 0 & 0 & u_{12} \\
        u_{21} \!\cos\!\eta & v_{22}e^{-i\delta} \!\sin\!\eta & v_{21} \!\sin\!\eta & u_{22} \!\cos\!\eta \\
        0 & v_{12}e^{-i\delta} & v_{11} & 0 \\
        -u_{21}\!\sin\!\eta & v_{22} e^{-i\delta} \!\cos\!\eta & v_{21} \!\cos\!\eta & -u_{22} \!\sin\!\eta 
    \end{array}
    \!\!\right)
\end{equation}
with respect to the computational basis $\{|00\rangle,|01\rangle,|10\rangle,|11\rangle\}$, where $u_{kl}$, $v_{kl}$ ($k,l=1,2$) and $\eta$ are real numbers with $\sum_k u_{kl}^2=\sum_l u_{kl}^2=\sum_k v_{kl}^2=\sum_l v_{kl}^2=1$.
%
%In other words, Alice's measurement basis is derived from two unitary operators $U = \sum_{k,l=1}^2 u_{kl} |\bar{k}\rangle\langle\bar{l}|$ and $V = \sum_{p,q=1}^2 v_{pq}\cdot e^{i\delta (q-1)} |\bar{p}\rangle\langle\bar{q}|$ acting on the subspaces of $\{|\bar{k}\rangle, |\bar{l}\rangle\} = \{ |00\rangle, |11\rangle \}$ and $\{|\bar{p}\rangle, |\bar{q}\rangle\} = \{|10\rangle, |01\rangle\}$, respectively. 
The unitary $D_{12}$ sends the initial state $|\phi\rangle_1 |\Phi\rangle_{23}$ to
\begin{equation}
  \begin{split}
    D_{12}^{\dagger} |\phi\rangle_1 |\Phi\rangle_{23} &= |00\rangle_{12}|\phi_{1+}\rangle_3 + |01\rangle_{12}|\phi_{2+}\rangle_3 \\
                                                      &+ |10\rangle_{12}|\phi_{1-}\rangle_3 + |11\rangle_{12}|\phi_{2-}\rangle_3.
  \end{split}
\end{equation}
After Alice measured qubits 1 and 2 using the computational basis, Bob's qubit 3 collapsed into the form of
$|\phi_{j\pm}\rangle_3 = \alpha |\phi_{\alpha}\rangle_3 + \beta |\phi_{\beta}\rangle_3$. 
This process is equivalent to performing a joint measurement of qubits 1 and 2 in the orthonormal basis $|\psi_j\pm\rangle_{12} = D_{12} |j'-1,j-1 \rangle $ with $j,j' = 1,2$.
%This process is equivalent to using the orthonormal measurement basis $|\psi_j\pm\rangle_{12}$ to measure qubits 1 and 2 jointly.
%Therefore, the design of the unitary operator $D_{12}$ is equivalent to construct Alice's measurement basis.
In other words, two of Alice's measurement states lie in the subspaces $\{|\bar{k}\rangle, |\bar{l}\rangle\} = \{ |00\rangle, |11\rangle \}$ and $\{|\bar{p}\rangle, |\bar{q}\rangle\} = \{|10\rangle, |01\rangle\}$, respectively.
The remaining two states are constructed as superpositions of the orthogonal states within the corresponding subspaces, i.e., the superposition of the state orthogonal to the first basis in $\{ |00\rangle, |11\rangle \}$ and the state orthogonal to the second basis in $\{|10\rangle, |01\rangle\}$.
Consequently, the parameters involved in constructing $D_{12}$ can be naturally associated with those of two unitary operators $U = \sum_{k,l=1}^2 u_{kl} |\bar{k}\rangle\langle\bar{l}|$ and $V = \sum_{p,q=1}^2 v_{pq}\cdot e^{i\delta (q-1)} |\bar{p}\rangle\langle\bar{q}|$ defined on the respective subspaces.

%The key to proving the perfect accomplishment of this protocol is whether conditions 
%(i) $\langle \phi_{\alpha} | \phi_{\alpha} \rangle = \langle \phi_{\beta} | \phi_{\beta} \rangle$
%and
%(ii) $\langle \phi_{\alpha} | \phi_{\beta} \rangle = \langle \phi_{\beta} | \phi_{\alpha} \rangle = 0$ are
%fulfilled, which yields four constraint equations in total.
%到目前为止，我们的方案因缺少某些先验条件（例如，施密特系数a_0和a_1的值）而包含了许多的未知参数。接下来我们将证明，我们的方案可以实现完美传态的前提是信道是最大纠缠的.
At this stage, our protocol contains multiple undetermined parameters due to the lack of certain prior conditions (for example, the values of Schmidt coefficients $a_{j=0,1}$).
We shall show that perfect teleportation in our scheme can only be achieved when the quantum channel is maximally entangled.
For the protocol to succeed perfectly, the following conditions must hold:
(i) $\langle \phi_{\alpha} | \phi_{\alpha} \rangle = \langle \phi_{\beta} | \phi_{\beta} \rangle$,
(ii) $\langle \phi_{\alpha} | \phi_{\beta} \rangle = \langle \phi_{\beta} | \phi_{\alpha} \rangle = 0$,
yielding four constraint equations in total.
By leveraging the unitarity of matrices $U$ and $V$, and the normalization condition $\sum_{j=0}^{1}a_j^2 =1$, we have
$a_0^2 = u_{12}^2 = v_{12}^2 = u_{21}^2 = v_{21}^2$ and
$a_1^2 = u_{11}^2 = v_{11}^2 = u_{22}^2 = v_{22}^2$.
%\begin{equation}  \label{parameters}
%  \begin{split}
%    a_0^2 &= u_{12}^2 = v_{12}^2 = u_{21}^2 = v_{21}^2, \\ 
%    a_1^2 &= u_{11}^2 = v_{11}^2 = u_{22}^2 = v_{22}^2.
%  \end{split}
%\end{equation}
Thus, the original four conditions simplify to the following form:
\begin{equation}  \label{CDTS} 
        a_0^2 u_{11}^2 \!=\! a_1^2 u_{12}^2, \ \ 
        (a_0^2 u_{21}^2 \!+\! a_1^2 u_{22}^2 \!\cdot\! e^{i\delta}) \sin\eta\cos\eta \!=\! 0,
\end{equation}
where we have assumed $u_{kl} = v_{kl}$ without loss of generality.

It is easy to show that the conditions hold if $a_0^2 u_{21}^2 + a_1^2 u_{22}^2 \cdot e^{i\delta} = 0$ or $\sin\eta\cos\eta = 0$.
If $a_0^2 u_{21}^2 + a_1^2 u_{22}^2 \cdot e^{i\delta} = 0$, one can obtain that $e^{i\delta} =-1$, $a_0^2 = a_1^2 =1/2$, and
\begin{subequations}  %\label{}
  \begin{equation}  \label{Am1}
    \begin{split}
       |\psi_{1+}\rangle_{12} &= \frac{1}{\sqrt{2}} (|00\rangle + |11\rangle)_{12}, \\
       |\psi_{1-}\rangle_{12} &= \frac{1}{\sqrt{2}} (|10\rangle - |01\rangle)_{12},
    \end{split}
  \end{equation}
  \begin{equation}  \label{Am2}
    \begin{split}
       |\psi_{2+}\rangle_{12} \!&=\! \frac{1}{\sqrt{2}} [\cos\eta  (-|00\rangle \!+\! |11\rangle) \!-\! \sin\eta (|10\rangle \!+\! |01\rangle)]_{12}, \\
       |\psi_{2-}\rangle_{12} \!&=\! \frac{1}{\sqrt{2}} [-\! \sin\eta (-|00\rangle \!+\! |11\rangle) \!-\! \cos\eta (|10\rangle \!+\! |01\rangle)]_{12}.
    \end{split}
  \end{equation}
\end{subequations}
By performing a unitary transformation $\mathcal{U}_1= |0\rangle_1 (\cos\eta \langle 0| + \sin\eta \langle 1|)_1 + |1\rangle_1 (-\sin\eta \langle 0| + \cos\eta \langle 1|)_1$, the maximally entangled states in Eq.~(\ref{Am2}) can be transformed as Bell bases $\frac{1}{\sqrt{2}} (|00\rangle - |11\rangle)_{12}$ and $\frac{1}{\sqrt{2}} (|10\rangle + |01\rangle)_{12}$.
%因此，这一新的方案与最初的方案是等价的。
Thus, this new scheme is equivalent to the standard one.
%The maximally entangled states in Eq.~(\ref{Am2}) are related to the Bell basis through a unitary transformation $\mathcal{U}_1= |0\rangle_1 (\cos\eta \langle 0| + \sin\eta \langle 1|)_1 + |1\rangle_1 (-\sin\eta \langle 0| + \cos\eta \langle 1|)_1$.
If $\sin\eta\cos\eta = 0$, one can obtain that $a_0^2 u_{12}^2 = a_1^2 u_{11}^2$, then the same conclusion holds.

\section{Two-qutrit channel} \label{ourscheme}

Extending the previous protocol to the two-qutrit channel follows naturally.
Similarly to the case of the two-qubit channel, the state that Alice can expect to deliver is still in the form of Eq.~(\ref{deliverstate}).

The transmission channel is described by
\begin{equation}  %\label{}
  |\Phi\rangle_{23} = a_0 |00\rangle_{23} + a_1 |11\rangle_{23} + a_2 |22\rangle_{23},
\end{equation}
where the Schmidt coefficients $a_{j=0,1,2}$ are real numbers satisfying $0 \leq a_j < 1$ and $\sum_{j=0}^2 a_j^2 = 1$.
The total state in the teleportation takes the form
\begin{equation}  \label{totalstate}
\begin{split}
  |\Psi\rangle_{123} &= |\phi\rangle_1 |\Phi\rangle_{23}  \\
                     &= [\alpha (a_0 |000\rangle + a_1 |011\rangle +a_2 |022\rangle) \\
                     &+ \beta (a_0 |100\rangle + a_1 |111\rangle +a_2 |122\rangle)]_{123}.
\end{split}
\end{equation}
%与two-qubit channel的例子相比， there are more possible combinations of the bases $\{|00\rangle, |01\rangle, |02\rangle, |10\rangle, |11\rangle, |12\rangle \}$
Compared with the two-qubit channel case, these basis vectors $\{|00\rangle, |01\rangle, |02\rangle, |10\rangle, |11\rangle, |12\rangle \}$ admit a richer variety of possible decompositions into mutually orthogonal subspaces.
%引入完美隐形传态的约束后， 依然存在三种维数相同的正交子空间的可能组合：$\{ |00\rangle, |01\rangle, |12\rangle \}$ and $\{ |10\rangle, |11\rangle, |02\rangle \}$， $\{ |00\rangle, |11\rangle, |02\rangle \}$ and $\{ |10\rangle, |01\rangle, |12\rangle \}$， 以及 $\{ |10\rangle, |01\rangle, |02\rangle \}$ and $\{ |00\rangle, |11\rangle, |12\rangle \}$.
After imposing perfect teleportation constraints, there remain three types of identity-dimensional orthogonal subspace decompositions: $\{ |00\rangle, |01\rangle, |12\rangle \}$ and $\{ |10\rangle, |11\rangle, |02\rangle \}$, $\{ |00\rangle, |11\rangle, |02\rangle \}$ and $\{ |10\rangle, |01\rangle, |12\rangle \}$, or $\{ |10\rangle, |01\rangle, |02\rangle \}$ and $\{ |00\rangle, |11\rangle, |12\rangle \}$.

%构造测量基的子空间的选取与施密特系数的大小关系相关。
The selection of subspaces for constructing the measurement bases is related to the size of the Schmidt coefficients.
If the Schmidt coefficients satisfy $\max \{a_j\}_{j=0,1,2} = a_1$, then Alice's measurements can be constructed via a unitary transformation acting on the subspace of $\{ |00\rangle, |11\rangle, |02\rangle \}$ and $\{ |10\rangle, |01\rangle, |12\rangle \}$. The unitary operators read as
\begin{equation*}
U \!=\! \left(\begin{array}{ccc}
      u_{11} & u_{12} & u_{13} \\
      u_{21} & u_{22} & u_{23} \\
      u_{31} & u_{32} & u_{33}
    \end{array}\right), \ \
V \!=\! \left(\begin{array}{ccc}
      v_{11} & v_{12} \!\cdot\! e^{i\delta_1} & v_{13} \!\cdot\! e^{i\delta_2} \\
      v_{21} & v_{22} \!\cdot\! e^{i\delta_1} & v_{23} \!\cdot\! e^{i\delta_2} \\
      v_{31} & v_{32} \!\cdot\! e^{i\delta_1} & v_{33} \!\cdot\! e^{i\delta_2}
    \end{array}\right),
\end{equation*}
where $u_{kl}$, $v_{kl}$, and $\delta_m$ ($k,l=1,2,3$, $m=1,2$) are real numbers.
%根据完美实现隐形传态的条件1和2，可直接推导出幺正矩阵元需满足的约束方程组。
The constraint equations for the unitary matrix elements can be directly derived from conditions (i) and (ii) for perfect quantum teleportation.
%需要指出的是，由于约束方程的数量远少于未知参数的数目，要具体分测量基的显示形式具有相当的难度。
Note that the number of constraint equations is significantly fewer than the number of unknown parameters, making it particularly challenging to determine the explicit form of the measurement bases analytically.
%因此，为简化分析，我们假定$u_{ij} = v_{ij}$。 进一步，我们限定$U$是一个SO（3）群中的旋转矩阵$R(\theta_1,\theta_2,\theta_3)$。
To simplify our analysis, we assume that $u_{kl} = v_{kl}$ and $U$ are restricted to be a rotation matrix $R(\theta_1,\theta_2,\theta_3)$ within the $SO(3)$ group.

For a two-qutrit channel, the unitary operator corresponding to Alice's joint measurement is of the form
\begin{widetext}
    \begin{equation}  \label{allbasis}
        \begin{split}
            D_{12} = \left(
            \begin{array}{cccccc}
                u_{11} & 0 & u_{13} & 0 & u_{12} & 0 \\
                0 & u_{12} e^{i\delta_1} & 0 & u_{11} & 0 & u_{13} e^{i\delta_2} \\
                u_{31} \cos\zeta & u_{32} e^{i\delta_1} \sin\zeta & u_{33} \cos\zeta & u_{31} \sin\zeta & u_{32} \cos\zeta & u_{33} e^{i\delta_2} \sin\zeta \\
                u_{21} & 0 & u_{23} & 0 & u_{22} & 0 \\
                0 & u_{22} e^{i\delta_1} & 0 & u_{21} & 0 & u_{23} e^{i\delta_2} \\
                -u_{31} \sin\zeta & u_{32} e^{i\delta_1} \cos\zeta & -u_{33} \sin\zeta & u_{31} \cos\zeta & -u_{32} \sin\zeta & u_{33} e^{i\delta_2} \cos\zeta 
            \end{array}
            \right)
        \end{split}
    \end{equation}
\end{widetext}
with respect to the computational basis $\{|00\rangle, |01\rangle, |02\rangle, |10\rangle, |11\rangle, |12\rangle\}$.
Similarly to the case of two-qubit channel, this unitary operation is equivalent to performing a joint measurement of qubits 1 and 2 in the orthonormal basis $|\psi_{j\pm} \rangle_{12} = D_{12} |j'-1,j-1 \rangle$ with $j'=1,2$ and $j=1,2,3$.
Among the three sets of measurement bases, the first two sets are chosen as states from the subspaces $\{ |00\rangle, |11\rangle, |02\rangle \}$ and $\{ |10\rangle, |01\rangle, |12\rangle \}$, while the remaining set is constructed analogously to the two-qubit channel case.
%%%%%%%%%%%%%%%%%%%%%
%\begin{widetext}
%\begin{equation}  \label{allbasis}
%  \begin{split}
%     |\psi_{1+}\rangle_{12} &= u_{11} |00\rangle + u_{12} |11\rangle + u_{13} |02\rangle, \ \ 
%     |\psi_{1-}\rangle_{12} = u_{21} |00\rangle + u_{22} |11\rangle + u_{23} |02\rangle, \\
%     |\psi_{2+}\rangle_{12} &= u_{11} |10\rangle + u_{12} \cdot e^{i\delta_1} |01\rangle + u_{13} \cdot e^{i\delta_2} |12\rangle, \ \ 
%     |\psi_{2-}\rangle_{12} = u_{21} |10\rangle + u_{22} \cdot e^{i\delta_1} |01\rangle + u_{23} \cdot e^{i\delta_2} |12\rangle, \\
%     |\psi_{3+}\rangle_{12} &= \cos\zeta (u_{31} |00\rangle + u_{32} |11\rangle + u_{33} |02\rangle) +\sin\zeta (u_{31} |10\rangle + u_{32} \cdot e^{i\delta_1} |01\rangle + u_{33} \cdot e^{i\delta_2} |12\rangle) \\
%                            &= u_{31} \cos\zeta |00\rangle + u_{32} \cdot e^{i\delta_1} \sin\zeta |01\rangle + u_{33} \cos\zeta |02\rangle + u_{31} \sin\zeta |10\rangle + u_{32} \cos\zeta |11\rangle + u_{33} \cdot e^{i\delta_2} \sin\zeta |12\rangle, \\
%     |\psi_{3-}\rangle_{12} &= -\sin\zeta (u_{31} |00\rangle + u_{32} |11\rangle + u_{33} |02\rangle) +\cos\zeta (u_{31} |10\rangle + u_{32} \cdot e^{i\delta_1} |01\rangle + u_{33} \cdot e^{i\delta_2} |12\rangle) \\
%                            &= -u_{31} \sin\zeta |00\rangle + u_{32} \cdot e^{i\delta_1} \cos\zeta |01\rangle - u_{33} \sin\zeta |02\rangle + u_{31} \cos\zeta |10\rangle - u_{32} \sin\zeta |11\rangle + u_{33} \cdot e^{i\delta_2} \cos\zeta |12\rangle.
%  \end{split}
%\end{equation}
%\end{widetext}
%%%%%%%%%%%%%%%%%%%%%
Perfect quantum teleportation requires $\sin^2\zeta = \cos^2\zeta = 1/2$, where we may take $\sin\zeta=\cos\zeta=1/\sqrt{2}$ by convention.
In this sense, after performing the above measurements on qubit 1 and qutrit 2, Alice's actions caused Bob's qutrit 3 to collapse into the state $|\phi_{j\pm}\rangle_3 = \alpha |\phi_{\alpha}\rangle_3 + \beta |\phi_{\beta}\rangle_3$, which are
    \begin{equation}
    \begin{split}
        |\phi_{1+}\rangle_3 &\!=\! \alpha (a_0 u_{11} \!|0\rangle \!+\! a_2 u_{13} |2\rangle)_3 \!+\! \beta a_1 u_{12} |1\rangle_3, \\
        |\phi_{1-}\rangle_3 &\!=\! \alpha (a_0 u_{21} \!|0\rangle \!+\! a_2 u_{23} |2\rangle)_3 \!+\! \beta a_1 u_{22} |1\rangle_3, \\
        |\phi_{2+}\rangle_3 &\!=\! \beta (a_0 u_{11} \!|0\rangle \!+\! a_2 u_{13} \!\cdot\! e^{-i\delta_2} \!|2\rangle)_3 \!+\! \alpha a_1 u_{12} \!\cdot\! e^{-i\delta_1} \!|1\rangle_3, \\
        |\phi_{2-}\rangle_3 &\!=\! \beta (a_0 u_{21} \!|0\rangle \!+\! a_2 u_{23} \!\cdot\! e^{-i\delta_2} \!|2\rangle)_3 \!+\! \alpha a_1 u_{22} \!\cdot\! e^{-i\delta_1} \!|1\rangle_3, \\
        |\phi_{3\pm}\rangle_3 &\!=\! \frac{1}{\sqrt{2}}[\alpha (\pm a_0 u_{31} |0\rangle \!+\! a_1 u_{32} \!\cdot\! e^{-i\delta_1}\! |1\rangle \!\pm\! a_2 u_{33} |2\rangle) \\
        & \!+\! \beta (a_0 u_{31} |0\rangle \!\pm\! a_1 u_{32} |1\rangle \!+\! a_2 u_{33} \!\cdot\! e^{-i\delta_2}\! |2\rangle)]_3.
    \end{split}
\end{equation}
The constraint equations for the unitary matrix elements and the Schmidt coefficients are given as
\begin{subequations}  \label{2Cdt}
  \begin{equation}  \label{2Cdt1}
  \begin{split}
    a_0^2 u_{11}^2 + a_2^2 u_{13}^2 &= a_1^2 u_{12}^2, \\ 
    a_0^2 u_{21}^2 + a_2^2 u_{23}^2 &= a_1^2 u_{22}^2,
  \end{split}
  \end{equation}
  \begin{equation}  \label{2Cdt2}
    a_0^2 u_{31}^2 + a_1^2 u_{32}^2 \cdot e^{i\delta_1} + a_2^2 u_{33}^2 \cdot e^{-i\delta_2} = 0.
  \end{equation}
\end{subequations}
Then the probabilities of Alice's outcome are given by the overlap $P_{j\pm} =_3\langle \phi_{j\pm}| \phi_{j\pm}\rangle_3$ as
\begin{equation}  \label{prob}
  \begin{split}
     P_{1+} &= P_{2+} = a_0^2 u_{11}^2 + a_2^2 u_{13}^2, \\ 
     P_{1-} &= P_{2-} = a_0^2 u_{21}^2 + a_2^2 u_{23}^2, \\
     P_{3\pm} &= \frac{1}{2} (a_0^2 u_{31}^2 + a_1^2 u_{32}^2 + a_2^2 u_{23}^2).
  \end{split}
\end{equation}

Next, one can easily prove that our scheme is sufficient and necessary for all quantum channels satisfying the perfect teleportation condition.\cite{PRA2004}
When $\max \{ a_j \}_{j=0,1,2} = a_1 \leq 1/\sqrt{2}$, Eqs.~(\ref{2Cdt1}) allow expressing two rotation angles (e.g., $\theta_1$ and $\theta_2$) in terms of the third ($\theta_3$), while constraining its admissible range.
This guarantees the existence of a solution for all valid channels.
In this sense, our scheme is necessary for any quantum channel satisfying $\max \{ a_j \}_{j=0,1,2} = a_1 \leq 1/\sqrt{2}$.
%Under the constraint (\ref{2Cdt1}), two out of the parameters $\theta_1$, $\theta_2$ and $\theta_3$ can be represented by the third one.
%In addition, the constraint (\ref{2Cdt1}) limits the range of the third parameter.
%Therefore, for any Schmidt coefficients that can achieve a perfect teleportation, our construction provides a valid scheme.
In contrast, consider a channel with $\max \{ a_j \}_{j=0,1,2} = a_1 = \sqrt{1/2 + \delta} > 1/\sqrt{2}$, ($\delta>0$).
From the unitary constraints $\sum_{k=1}^3 u_{kl}^2 = \sum_{l=1}^3 u_{kl}^2 = 1$, one can derive $a_0^2 u_{31}^2 + a_2^2 u_{33}^2 - a_1^2 u_{32}^2 = a_0^2 + a_2^2 - a_1^2 < 0$.
%This inequality implies that our scheme cannot provide a set of measurement bases for any quantum channel not satisfying the perfect teleportation condition.
This implies that our scheme is sufficient for any quantum channel satisfying $\max \{ a_j \}_{j=0,1,2} = a_1 \leq 1/\sqrt{2}$.
%%%%%%%%%This inequality precludes the existence of real solutions for the matrix elements, confirming our scheme's incompatibility with non-compliant channels.
In particular, this analysis can be extended symmetrically to cases where $a_0$ or $a_2$ are maximal, thus exhaustively characterizing all perfect teleportation-capable channels.
This completes the proof.

\section{Entanglement and classical information}

%一个完美隐形传态方案有三个关键因素，建立量子信道，爱丽丝进行的联合测量，以及向鲍勃传递经典信息，它们均需小号一定的资源。
A perfect quantum teleportation scheme involves three key elements: establishing a quantum channel, the joint measurement performed by Alice, and the transmission of classical information to Bob. Each of these steps requires the consumption of certain resources.
Our scheme provides a continuous region to explore these resources in the perfect teleportation of a qubit, where the Schmidt coefficients of the bipartite states can be selected at random.
%此外，通过上述三个关键因素，我们还比较了我们方案与参考文献PRA2004的方案。
Also, by examining the three key elements mentioned above, we have compared our scheme with the one presented in Ref.~\cite{PRA2004}.

The quantifiers of the above three resources have been mentioned in Ref.~\cite{chen2022perfect}, which are
\begin{subequations}  %\label{}
  \begin{equation}  \label{qtt1}
    \mathcal{E} (|\Phi\rangle_{23}) = -\sum_{i=0}^{2} a_i^2 \log_2 a_i^2,
  \end{equation}
  \begin{equation}  \label{qtt2}
    \mathcal{E}_{12} = \sum_{j=1}^{3} [P_{j+} \mathcal{E} (|\psi_{j+}\rangle_{12}) + P_{j-} \mathcal{E} (|\psi_{j-}\rangle_{12})],
  \end{equation}
  \begin{equation}  \label{qtt3}
    \mathcal{H}_{12} = -\sum_{j=1}^{3} (P_{j+} \log_2 P_{j+} + P_{j-} \log_2 P_{j-}),
  \end{equation}
\end{subequations}
where
$\mathcal{E} (|\Phi\rangle_{23})$ is the entanglement entropy of the quantum channel,
$\mathcal{E}_{12}$ is the entanglement of Alice's joint measurement defined by the average of the bases, 
$\mathcal{H}_{12}$ is the Shannon entropy of the distribution (\ref{prob}).
It is worth noting that (\ref{qtt2}) can be derived by direct generalization in the work of Li et al.~\cite{li2000probabilistic}.
The entanglement of the qubit-qutrit states (Eq.~(\ref{allbasis})) can be presented as a monotone increasing function of the concurrence, which are
\begin{equation}  %\label{}
 \begin{split}
    \mathcal{C} (|\psi_{1+}\rangle_{12}) &\!=\! \mathcal{C} (|\psi_{2+}\rangle_{12}) \!=\! 4 u_{12}^2 (u_{11}^2 \!+\! u_{13}^2), \\
    \mathcal{C} (|\psi_{1-}\rangle_{12}) &\!=\! \mathcal{C} (|\psi_{2-}\rangle_{12}) \!=\! 4 u_{22}^2 (u_{21}^2 \!+\! u_{23}^2), \\
    \mathcal{C} (|\psi_{3\pm}\rangle_{12}) &\!=\! 2u_{31}^2 u_{32}^2 (1 \!-\! \cos\delta_1) \!+\! 2u_{31}^2 u_{33}^2 (1\!-\! \\
    & \cos\delta_2) +\! 2u_{32}^2 u_{33}^2 [1 \!-\! \cos(\delta_1\!+\!\delta_2)], \\
 \end{split}
\end{equation}
where
\begin{equation*}
    \begin{split}
        &\cos\delta_1 = -\frac{ a_0^4 u_{31}^4 + a_1^4 u_{32}^4 - a_2^4 u_{33}^4 }{2a_0^2 a_1^2 u_{31}^2 u_{32}^2}, \\
        &\cos\delta_2 = -\frac{ a_0^4 u_{31}^4 + a_2^4 u_{33}^4 - a_1^4 u_{32}^4 }{2a_0^2 a_2^2 u_{31}^2 u_{33}^2}, \\
        &\cos(\delta_1 + \delta_2) = -\frac{ a_1^4 u_{32}^4 + a_2^4 u_{33}^4 - a_0^4 u_{31}^4 }{2a_1^2 a_2^2 u_{32}^2 u_{33}^2}.
    \end{split}
\end{equation*}
%下面我们将列出量子信道的两种特殊情况来更清晰地分析这三个量的性质。
Next, we will list two special cases of quantum channels to clarify the properties of these three quantities.

{\bf Case I:} $a_0 = a_1$ or $a_2 = a_1$.
Since the constraints are symmetric under the exchange of $a_0$ and $a_2$, it is sufficient to analyze the case $a_2 = a_1$ without loss of generality.
The scheme in Ref.~\cite{PRA2004} for this case can give an analytical expression of the entanglement of Alice's joint measurement, that is
\begin{equation}  \label{pra20041}
  \mathcal{E}_{12}' \!\!=\!\! H\! \left(\! \frac{1 \!\!+\!\! \sqrt{\frac{1}{3} \!\!+\!\! \frac{2}{9} [ \cos\xi_1 \!\!+\!\! \cos\xi_2 \!\!+\!\! \cos(\xi_1 \!\!-\!\! \xi_2) ]}}{2} \!\right)\!,
\end{equation}
where $H(x) = -x \log_2 x - (1-x) \log_2 (1-x)$ is the binary entropy with $\xi_1 = \pi - \arccos \frac{2a_1^4 - a_2^4}{2 a_1^4}$ and $\xi_2 = \pi + \arccos \frac{a_2^2}{2 a_1^2}$.
In contrast, our scheme involves more degrees of freedom ($\{\theta_j\}_{j=1,2,3}$, $\{\delta_j\}_{j=1,2}$). The constraint $a_2 = a_1$ only partially restricts these parameters, necessitating a numerical exploration of the accessible entanglement regimes.

\begin{figure}[htbp]
  \centering
    \subfigure{
      \includegraphics[width=0.4\textwidth]{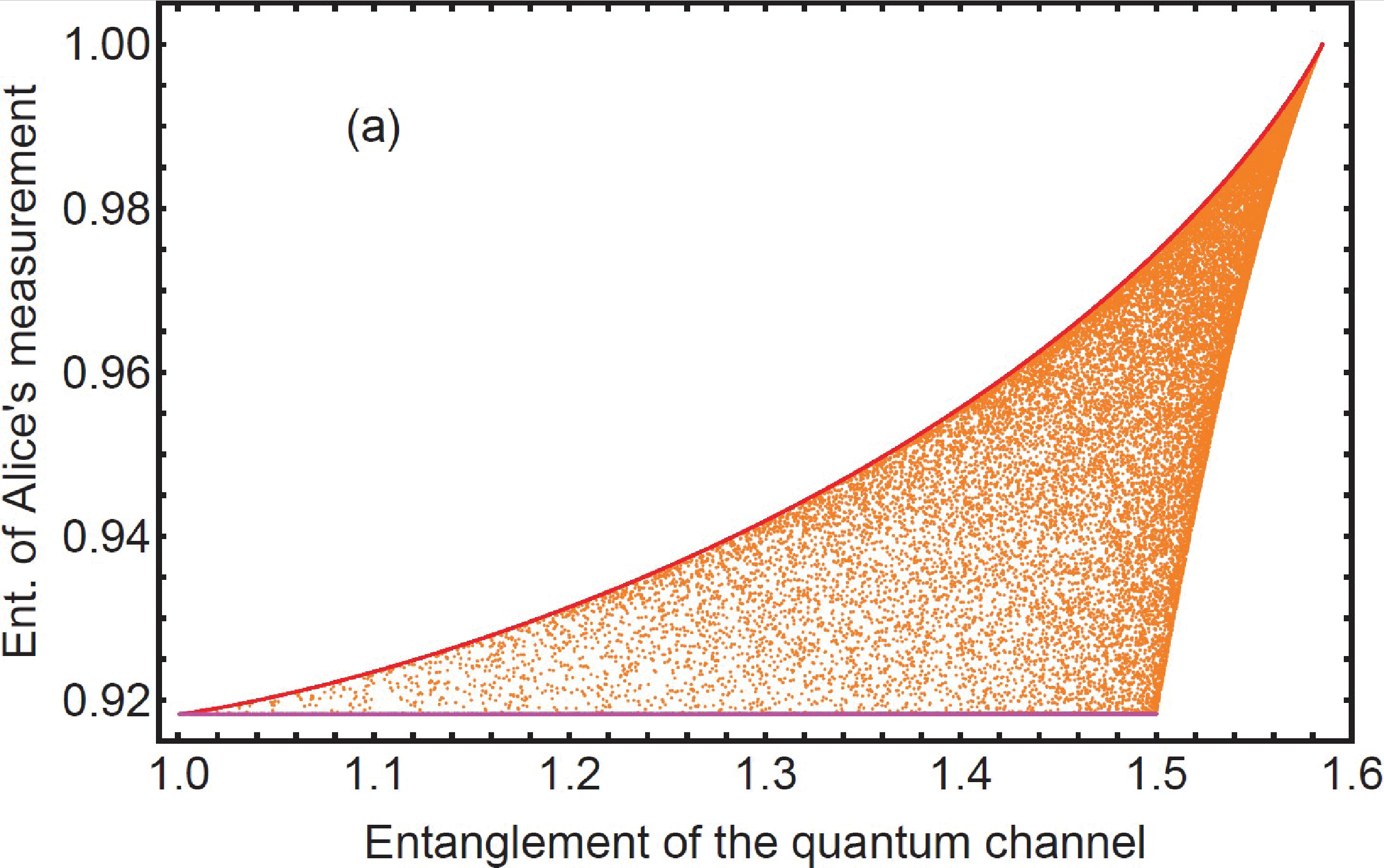}
      } \\
    \subfigure{
      \includegraphics[width=0.4\textwidth]{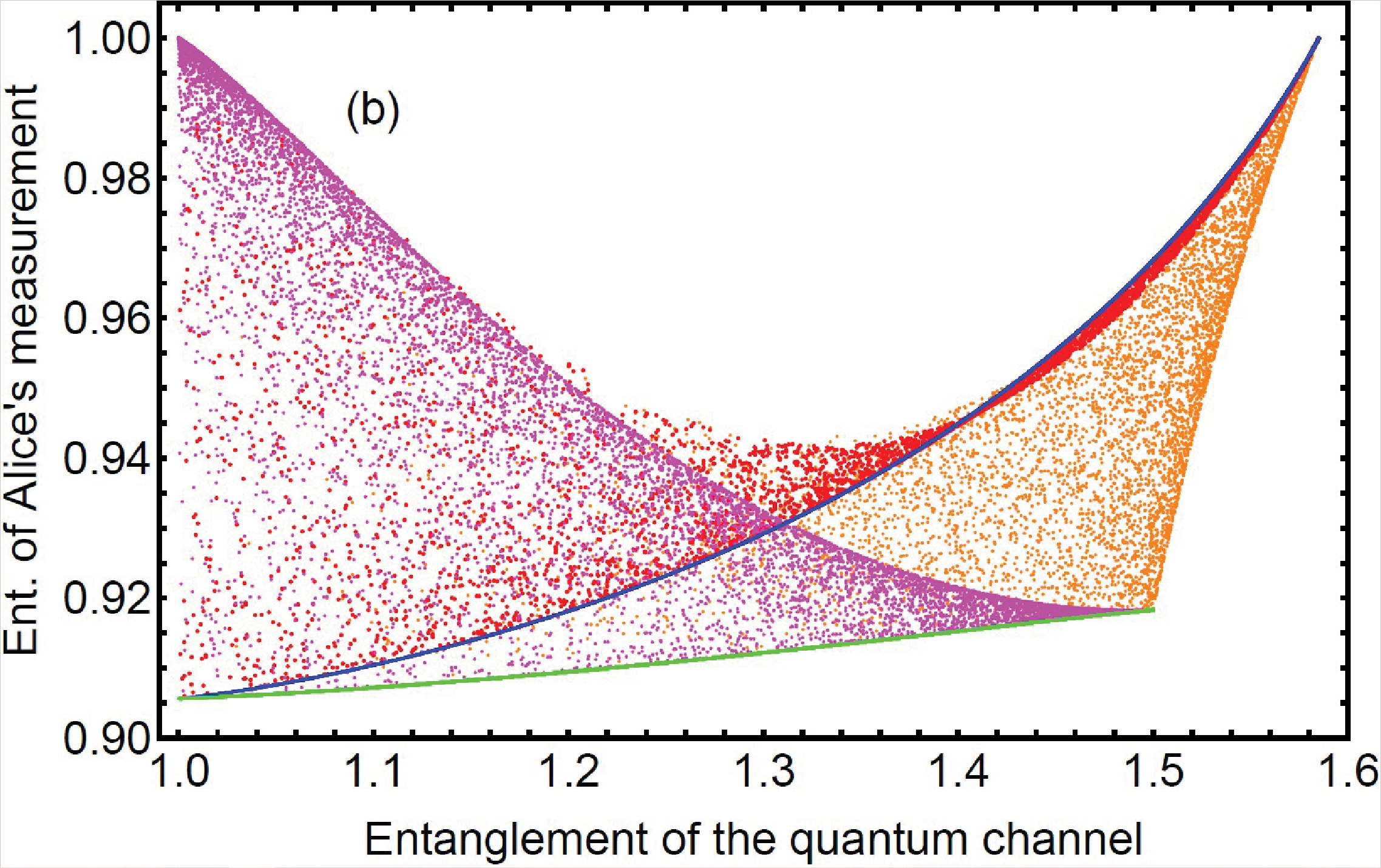}
      } 
  \caption{
Entanglement degree regimes for Alice's joint measurement and the quantum channel.
(a) Results from Ref.~\cite{PRA2004} scheme: red curve (case I) and magenta curve (case II).
(b) Our proposed scheme (Section \ref{ourscheme}): red region (case I) and magenta region (case II).
In addition, more specific parameter configurations are shown in (b). 
The blue curve shows case I with $\theta_2 = \pi/4$, $\theta_3 \in [0,\pi/4]$ and $\theta_1 =\frac{1}{2} \arctan(-\sqrt{2}\cot2\theta_3)$.
The green curve shows case II with $\theta_1 = \pi/4$, $\theta_2 = 0$ and $\theta_3 \in [\arcsin\sqrt{1/3}, \pi/4]$.
}  \label{Fig2}
\end{figure}

{\bf Case II:} $a_1 = 1/\sqrt{2}$.
%类似地，文献\cite{PRA2004}中的方案在该情况下的Alice的联合测量的纠缠度可表示为
For this configuration, Ref.~\cite{PRA2004} yields the entanglement measure
\begin{equation}  \label{pra20042}
  \mathcal{E}_{12}' \!\!=\!\! H\! \left(\! \frac{1 \!\!+\!\! \sqrt{\frac{1}{3} \!\!+\!\! \frac{2}{9} [ \cos\xi_1' \!\!+\!\! \cos\xi_2' \!\!+\!\! \cos(\xi_1' \!\!-\!\! \xi_2') ]}}{2} \!\right)\!
\end{equation}
with $\xi_1' = \pi - \arccos \frac{ a_0^4 + 1/4 - a_2^4}{a_0^2}$ and $\xi_2' = \pi + \arccos \frac{a_0^4 + a_2^4 - 1/4}{2 a_0^2 a_2^2}$.
In our scheme, although constrained by Eqs.~(\ref{2Cdt}) and the condition $a_1 = 1/\sqrt{2}$, the parameter space still cannot fully determine the relationship between parameters $\{\theta_j\}_{j=1,2,3}$ and $\{\delta_j\}_{j=1,2}$, and the Schmidt coefficients $\{ a_j \}_{j=0,1,2}$.
Consequently, numerical methods are required to determine the achievable entanglement regimes between Alice's joint measurement and the quantum channel.

%图1比较了不同方案下Alice联合测量与量子信道的纠缠度关系：(a)基于文献\cite{PRA2004}方案的曲线结果，(b)采用本文方案（第\ref{ourscheme}节）的参数空间区域分布。
Figure~\ref{Fig2} compares the entanglement degree relationships between Alice's joint measurement and the quantum channel: (a) results from Ref.~\cite{PRA2004} protocol, (b) our scheme's results (Sect.~\ref{ourscheme}).
%容易发现，Alice联合测量的纠缠度在量子信道最大纠缠时达到最大值1，当任一施密特系数$a_{j=0,1,2}$趋于零时则逼近最小值。
It is easy to find that the entanglement of Alice's joint measurement reaches its maximum when the quantum channel is maximally entangled, and approaches a minimum as any Schmidt coefficient $a_{j=0,1,2}$ vanishes.
%显然，我们的方案具有更低的下界。通过可分析的表达式，当任意施密特系数趋于零时，我们的方案得到的Alice联合测量的纠缠度为$\mathcal{E}_{12} = 1/2 + 1/2 H(3/4) \approx 0.906$，而\cite{PRA2004}的方案得到的值为$\mathcal{E}_{12}' = H(2/3) \approx 0.918$。
In particular, our scheme achieves a superior lower bound. When any $a_j \rightarrow 0$, the entanglement degree of Alice's joint measurement yields $\mathcal{E}_{12} = \frac{1}{2} + \frac{1}{2} H(\frac{3}{4}) \approx 0.906$ for our protocol, compared to $\mathcal{E}_{12}' = H(\frac{2}{3}) \approx 0.918$ in Ref.~\cite{PRA2004}.

\begin{figure}[htbp]
  \centering
  \includegraphics[width=0.4\textwidth]{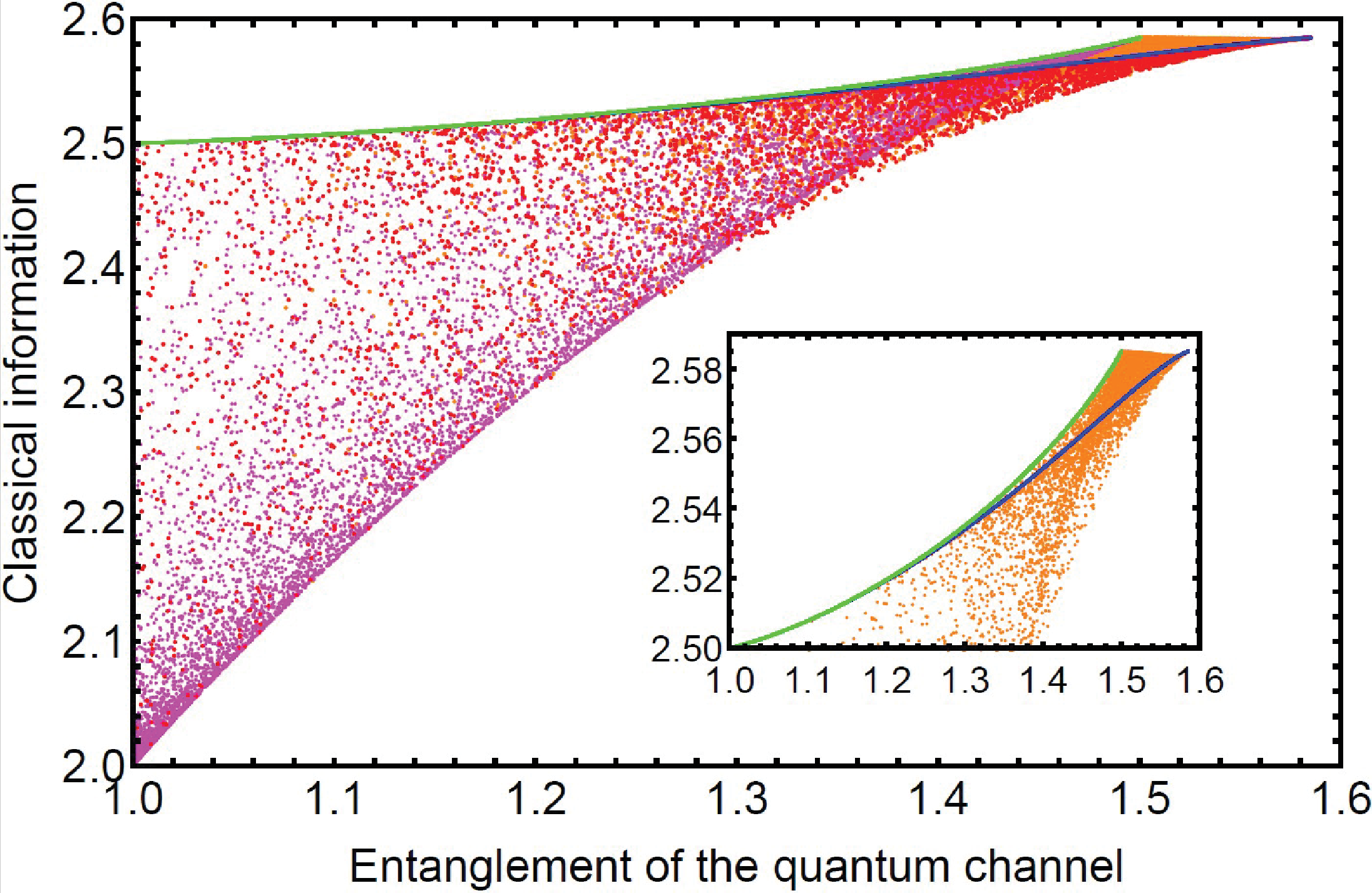}
  \caption{
The classical bits sent to Bob vs. the entanglement of the quantum channel.
The regions and curves of different colors correspond one-to-one with the various cases in Fig.~1b.
  }  \label{Fig3}
\end{figure}

One can also note that when the quantum channel entanglement is 1, the scheme in Ref.~\cite{PRA2004} is constrained to a unique protocol of Alice's joint measurement basis due to its fundamental design (omitting a common phase factor). In contrast, our scheme provides flexible non-unique protocols (detailed examples are shown in the next section).
Additionally, the region for case I in our scheme (red region in panel (b)) shows a non-monotonic relationship between the entanglement of Alice's joint measurement and the quantum channel.
This demonstrates that the entanglement of Alice's joint measurement does not represent Alice's ability to utilize the quantum channel.

Figure~\ref{Fig3} shows the regions of the classical bits sent to Bob versus the entanglement of the quantum channel, along with the specific regions and curves for cases I and II.
%在{PRA2004}的方案中，Alice得到的结果的概率与信道无关，而我们的方案可以明显观察到classical communication cost 和量子信道的纠缠之间存在overall upward trend。
In the scheme of Ref.~\cite{PRA2004}, the probabilities of Alice's outcome remain independent of the quantum channel, whereas our scheme demonstrates an overall upward trend between the classical communication cost and the entanglement of the quantum channel.
When the quantum channel reduces to a two-qubit Bell state, the minimum number of classical bits required reaches its lower bound of 2 bits.
This implies that our scheme contains the case of the two-qubit channel naturally.
We will show this in the next section.
%Moreover, as shown in panel (b) of Fig.~\ref{Fig2} and Fig.~\ref{Fig3}, the greater the entanglement resource consumption in Alice's joint measurements, the lower the classical communication costs.
%The minimum of classical bits can be found at the limit of $a_0 \rightarrow 0$ with the parameters chosen as $\theta_1 = 0$ or $\pi/2$, $\theta_2 = 0$ or $\pi/2$, and $\theta_3 = \pi/4$.

%我们观察到，Alice测量的平均纠缠度与发送给Bob的经典信息的Shannon熵之和呈现出一种折衷关系：Alice联合测量中纠缠资源消耗越大，经典通信成本越低。
We observe that the sum of the average entanglement of Alice's measurement and the Shannon entropy of the classical information sent to Bob exhibits a trade-off relation: the greater the entanglement resource consumption in Alice's joint measurement, the lower the classical communication costs.
By Eqs.~(\ref{qtt2}) and (\ref{qtt3}), the upper bound of $\mathcal{E}_{12} + \mathcal{H}_{12}$ can be easily obtained as
\begin{equation}
\begin{split}
    &\mathcal{E}_{12} + \mathcal{H}_{12}  \\
    &\leq \sum_{j=1}^3 \{ P_{j+}[H(\frac{1+\sqrt{1 - C_{j+}}}{2}) + \log_2 P_{j+}]  \\
    &+ P_{j-}[H(\frac{1 + \sqrt{1 - C_{j-}}}{2})+\log_2 P_{j-}] \}
\end{split}
\end{equation}
with $C_{1+} = C_{2+} = 1-\cos^4\theta_1 \sin^4\theta_3$, $C_{1-} =  C_{2-} = 1-\sin^4\theta_1 \sin^4\theta_3$, $C_{3\pm} = \cos^2\theta_3 (1 + \sin^2\theta_3)$, and $P_{1+} = P_{2+} = \frac{\sin^2\theta_1 + \cos^2\theta_1 \cos^2\theta_3}{4+2\cos2\theta_3}$, $P_{1-} = P_{2-} = \frac{\cos^2\theta_3 + \cos^2\theta_1 \sin^2\theta_3}{4+2\cos2\theta_3}$, $P_{3\pm} = \frac{\cos^2\theta_3}{4+2\cos2\theta_3}$, where $\theta_1 =\frac{1}{2} \arctan(-\sqrt{2}\cot2\theta_3)$, and $\theta_3 = \frac{1}{2} \arccos (\frac{1-2a_1^2}{a_1^2})$.
%对于一个固定的量子信道，$\mathcal{E}_{12} + \mathcal{H}_{12}$的上界由$a_1$的一个函数所bound。
For a fixed quantum channel, the quantity $\mathcal{E}_{12} + \mathcal{H}_{12}$ is upper-bounded by a function of the Schmidt coefficient $a_1$.

%量$\mathcal{E}_{12} + \mathcal{H}_{12}$的下界更重要，因为它的下界体现出了Alice利用固定的信道传态所必须付出的最小代价。
%The more essential inequality of the quantity $\mathcal{E}_{12} + \mathcal{H}_{12}$ is its lower bound, which presents the minimal costs that Alice uses a quantum channel to teleport the target state.
A more physically significant bound is the lower bound on $\mathcal{E}_{12} + \mathcal{H}_{12}$, characterizing the minimal teleportation cost that Alice incurs when using the fixed quantum channel.
%$\mathcal{E}_{12} + \mathcal{H}_{12}$的下界是一个与施密特系数相关的分段函数
This lower bound can be characterized as a piecewise function of the entanglement entropy of the quantum channel $\mathcal{E}(|\Phi\rangle_{23}) = \mathcal{E}$:
\begin{equation}
    \mathcal{E}_{12} + \mathcal{H}_{12} \geq 
    \begin{cases}
     f_1(\mathcal{E}),\ \  \mathcal{E} \in (1,\frac{3}{2}) \\
     f_2(\mathcal{E}),\ \  \mathcal{E} \in (\frac{3}{2},\log_2 3)
    \end{cases}\ ,
\end{equation}
where
\begin{equation*}
    \begin{split}
       & f_1(\mathcal{E}) = g (q) = \frac{q+3}{q+1} + 2\log_2 (q+1) - \\
       & \frac{2q+1}{(q+1)^2} \log_2 (2q+1) - \frac{q(2q+1)}{(q+1)^2} \log_2 q,\\
       & f_2(\mathcal{E}) = k\mathcal{E} + b.
    \end{split}
\end{equation*}
Here, the parameter $q$ exhibits a positive correlation with the entanglement entropy of the quantum channel $\mathcal{E}$, obeying the relation $H(q) = 2 (\mathcal{E} - 1)$, and $k=\frac{5/3-\log_2 3}{\log_2 3 -3/2}$, $b = 2\log_2 3 +\frac{11}{6} -\frac{1}{4\log_2 3 -6}$.

\section{Specific cases}

%本节我们将通过一个特定案例的定量分析来说明我们的方案在信道退化为二维情形时协议的不唯一性。
In this section, we will demonstrate through quantitative analysis of a specific case that our protocol exhibits non-uniqueness when the two-qutrit channel degenerates to a two-qubit scenario.
%注意到在该极限情形下我们的方案在case I 和 II下测量的纠缠有着相同的区域，简单起见我们只考虑case I。
Remarkably, Cases I and II produce identical entanglement measurement regions in this limit. We therefore focus on Case I to streamline our discussion.

%量子信道的维度退化体现为任一施密特系数为0，在此我们假设case I中$a_0 = 0$.
The degeneracy of channel dimensions emerges when any Schmidt coefficient approaches zero.
%作为一个典型的例子，我们分析了条件$a\u 0=0$下的情形I，其中two-qutrit信道的Schmidt分解中的第一项消失。
As a prototypical example, we analyze case I under the condition $a_0 = 0$, where the first term in the Schmidt decomposition of the two-qutrit channel vanishes.
%To characterize the specific cases in case I, one can consider $a_0 = 0$ for case I, without loss of generality.
%Without loss of generality, we consider the case of $a_0 = 0$.
Under this condition, Alice's joint measurement basis reduces to
\begin{equation}  \label{tht1}
  \begin{split}
     |\psi_{1+}\rangle_{12} &= \cos \theta_1 |00\rangle + \frac{\sin \theta_1}{\sqrt{2}} (-|11\rangle + |02\rangle ), \\ 
     |\psi_{1-}\rangle_{12} &= \sin \theta_1 |00\rangle + \frac{\cos \theta_1}{\sqrt{2}} (|11\rangle - |02\rangle), \\
     |\psi_{2+}\rangle_{12} &= \cos \theta_1 |10\rangle - \frac{\sin \theta_1}{\sqrt{2}} ( |01\rangle + |12\rangle ), \\ 
     |\psi_{2-}\rangle_{12} &= \sin \theta_1 |10\rangle +\frac{\cos \theta_1}{\sqrt{2}} ( |01\rangle + |12\rangle ), \\
     |\psi_{3\pm}\rangle_{12} &= \frac{1}{2} [(|11\rangle + |02\rangle) \pm (|01\rangle - |12\rangle)]
  \end{split}
\end{equation}
with $\theta_1 \in [0,\pi/2]$, or
\begin{equation}  \label{tht2}
    \begin{split}
     |\psi_{1+}\rangle_{12} &= \cos \theta_2 |00\rangle +\frac{\sin \theta_2}{\sqrt{2}} ( |11\rangle + |02\rangle ), \\ |\psi_{1-}\rangle_{12} &= \frac{1}{\sqrt{2}} (|11\rangle + |02\rangle), \\
     |\psi_{2+}\rangle_{12} &= \cos \theta_2 |10\rangle +\frac{\sin \theta_2}{\sqrt{2}} ( |01\rangle - |12\rangle ), \\ |\psi_{2-}\rangle_{12} &= \frac{1}{\sqrt{2}} (|01\rangle - |12\rangle), \\
     |\psi_{3\pm}\rangle_{12} &= [-\frac{\sin \theta_2}{\sqrt{2}}|00\rangle + \frac{\cos \theta_2}{2} (|11\rangle + |02\rangle)] \\
     &\pm [-\frac{\sin \theta_2}{\sqrt{2}}|10\rangle + \frac{\cos \theta_2}{2} (|01\rangle - |12\rangle)]
  \end{split}
\end{equation}
with $\theta_2 \in [0,\pi/2]$.
%由于以上两组测量基不能通过局域幺正变换互相转化，因此通过我们的方案可以很容易得到不唯一的teleportation protocols。
Since the above two sets of measurement bases cannot be transformed into each other through a local unitary transformation, it is easy to get non-unique teleportation protocols through our scheme.
%Additionally, when $\theta_2 = 0$ or $\pi/2$, or $\theta_1 = 0$ or $\pi/2$, $\mathcal{E}_{12}$ reaches its maximum value of 1, while for $\theta_2 = \pi/4$, or $\theta_1 = \pi/4$, $\mathcal{E}_{12}$ attains its minimum value of approximately 0.906.

For simplicity, we assume that Alice performs her measurements in the form of Eq.~\ref{tht1}. Then the probabilities of Alice's outcome and the entanglement of Alice's joint measurement can be given by
\begin{subequations}  \label{Probtht1}
    \begin{equation}
    \mathcal{P}_{1+} \!\!=\!\! \mathcal{P}_{2+} \!\!=\!\! \frac{\sin^2{\theta_1}}{4}, \ 
    \mathcal{P}_{1-} \!\!=\!\! \mathcal{P}_{2-} \!\!=\!\! \frac{\cos^2{\theta_1}}{4}, \ 
    \mathcal{P}_{3\pm} \!\!=\!\! \frac{1}{4},
\end{equation}
\begin{equation}
    \mathcal{E}_{12} \!\!=\!\! \frac{1}{2}\!\! \left[ \! 1 \!\!+\!\! \sin^2\!{\theta_1} \!H\! (\!\frac{1 \!\!+\!\! \cos^2\!{\theta_1}}{2}\!) \!\!+\!\! \cos^2\!{\theta_1} \!H\! (\!\frac{1 \!\!+\!\! \sin^2\!{\theta_1}}{2}\!)\! \right]\!,
\end{equation}
\end{subequations}
where $H(x)$ is the binary entropy mentioned above.
Further, the Shannon entropy of the distribution (\ref{Probtht1}) is given by
\begin{equation}
    \mathcal{H}_{12} \!\!=\!\! 2 \!\!-\!\! \frac{1}{2} (\sin^2{\theta_1} \log_2{\sin^2{\theta_1}} \!\!+\!\! \cos^2{\theta_1} \log_2{\cos^2{\theta_1}}).
\end{equation}

\begin{figure}[htbp]
 \centering
  \includegraphics[width=0.4\textwidth]{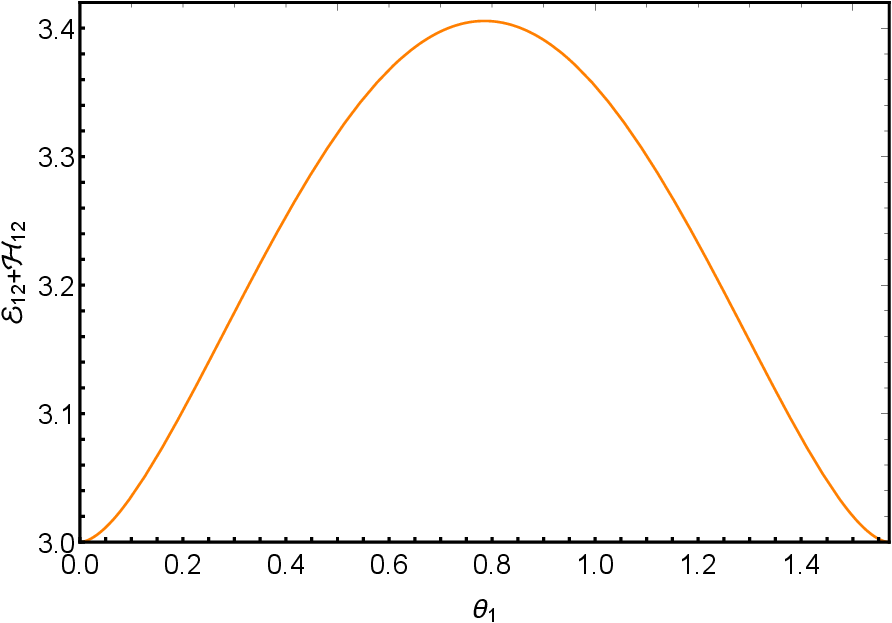}
  \caption{
The relationship between the total resources that Alice costs and the parameter $\theta_1$ when $a_0 = 0$.
  }  \label{Fig4}
\end{figure}

Figure~\ref{Fig4} shows the relationship between the total resources that Alice costs ($\mathcal{E}_{12} + \mathcal{H}_{12}$) and the parameter $\theta_1$. 
The minimum of $\mathcal{E}_{12} + \mathcal{H}_{12}$ occurs when $\theta_1 = 0$ or $\pi/2$.
In this scenario, the measurement of qubit 1 and qutrit 2 in bases $|\psi_{1+}\rangle_{12}$ and $|\psi_{2+}\rangle_{12}$ (with $\theta_1 = 0$) or $|\psi_{1-}\rangle_{12}$ and $|\psi_{2-}\rangle_{12}$ (with $\theta_1 = \pi/2$) does not collapse qutrit 3 into the form of $\alpha|\phi_{\alpha}\rangle + \beta|\phi_{\beta}\rangle$.
This implies that our scheme reduces to two-qubit Bell state teleportation.
Therefore, the entanglement of Alice's measurement and the Shannon entropy of the distribution are $\mathcal{E}_{12} = 1$ and $\mathcal{H}_{12} = 2$, respectively.
%the sum of the entanglement of Alice's joint measurement and the classical information sent to Bob 
In addition, the maximum of $\mathcal{E}_{12} + \mathcal{H}_{12}$ occurs when $\theta_1 = \pi/4$.
In this case, the measurement collapses the initial state $|\phi\rangle_1 |\Phi\rangle_{23}$ as
\begin{equation}
\begin{split}
    |\phi_{1\pm}\rangle_3 &= \pm \frac{1}{2\sqrt{2}} (\alpha |2\rangle - \beta |1\rangle)_3, \\
    |\phi_{2\pm}\rangle_3 &= \mp \frac{1}{2\sqrt{2}} (\alpha |1\rangle + \beta |2\rangle)_3, \\
    |\phi_{3\pm}\rangle_3 &= \frac{1}{2\sqrt{2}} [\alpha (|1\rangle \pm |2\rangle) + \beta (\pm |1\rangle - |2\rangle)]_3
\end{split}
\end{equation}
with the probabilities $\mathcal{P}_{1\pm} = \mathcal{P}_{2\pm} = 1/8$ and $\mathcal{P}_{3\pm} = 1/4$.
The entanglement of Alice's measurement and the Shannon entropy of the distribution are $\mathcal{E}_{12} = \frac{1}{2} [1 + H(\frac{3}{4})] \approx 0.906$ and $\mathcal{H}_{12} = 2 + \frac{1}{2} \log_2 \frac{1}{2} = \frac{5}{2}$.
%The minimum of $\mathcal{E}_{12} + \mathcal{H}_{12}$ occurs when $\theta_1 = 0$ or $\pi/2$. 
%It is easy to find that only four measurement bases are valid if $\theta_1 = 0$ ($|\psi_{1-}\rangle$, $|\psi_{2-}\rangle$, $|\psi_{3\pm}\rangle$) or $\theta_1 = \pi/2$ ($|\psi_{1+}\rangle$, $|\psi_{2+}\rangle$, $|\psi_{3\pm}\rangle$).
%These results demonstrate that the entanglement in the measurement and the classical bits sent to Bob exist in a trade-off relation.

%\begin{equation*}
%    \begin{split}
%        C_{1+} &= C_{2+} = (\cos^2\theta_3 + \sin^2\theta_1 \sin^2\theta_3)\times \\
%        &[2-(\cos^2\theta_3 + \sin^2\theta_1 \sin^2\theta_3)], \\
%        C_{1-} &=  C_{2-} = (\cos^2\theta_1 + \sin^2\theta_1 \cos^2\theta_3)\times \\
%        &[2-(\cos^2\theta_1 + \sin^2\theta_1 \cos^2\theta_3)], \\
%        C_{3\pm} &= \cos^2\theta_3 (1 + \sin^2\theta_3),
%    \end{split}
%\end{equation*}
%and
%\begin{equation*}
%    \begin{split}
%        P_{1+} &= P_{2+} = \frac{\sin^2\theta_1 + \cos^2\theta_1 \cos^2\theta_3}{4+2\cos2\theta_3}, \\
%        P_{1-} &= P_{2-} = \frac{\cos^2\theta_3 + \cos^2\theta_1 \sin^2\theta_3}{4+2\cos2\theta_3}, \\
%        P_{3\pm} &= \frac{\cos^2\theta_3}{4+2\cos2\theta_3}.
%    \end{split}
%\end{equation*}

\section{summary}

%通过推广原始的two-qubit信道的teleportation的一种等价方案，我们得到了一个利用部分纠缠的two-qutrit信道完美teleportation的方案。
By extending an equivalent scheme for the original teleportation protocol using the two-qubit channel, we propose a scheme for perfect teleportation utilizing the partially entangled two-qutrit channel.
%对任意的满足完美隐形传态的two-qutrit部分纠缠信道，利用我们的方案提出一个隐形传态协议是充分必要的。
For any partially entangled two-qutrit channel that satisfies the perfect teleportation condition, it is sufficient and necessary to propose a teleportation protocol using our scheme.
%与文献【PRA2004】的方案相比，我们的方案在相同的信道下会消耗更少的经典和量子资源。而与文献【PRA2022】的方案相比，我们的方案适用于更广的信道范围。
Compared with the scheme in~\cite{PRA2004}, our scheme consumes fewer classical and quantum resources for the same channel.
In addition, our scheme is suitable for a wider channel range than the scheme in~\cite{chen2022perfect}.
%我们还建立了Alice测量纠缠度与发送给Bob的经典信息之间的折衷关系，从而量化了这两个量的下界。
We also establish a trade-off relation between the entanglement of Alice's measurement and the classical information sent to Bob, which quantifies the lower bound of these two quantities.

%值得注意的是，当推广到更高维信道的情况时，我们的方案就会失效。
%It is worth noting that our scheme is invalid for extending to higher dimensional channels.
%一个直接的原因是，随着Alice手中系统的维度$d_a$的增加，可计算基分解成相互正交的子空间的组合数呈指数上升（$2^{d_a/2-1}-1$）。这大大提高了构造Alice的测量基的复杂度。
%因此，一个自然的问题是，是否存在一类隐形传态协议，其高维情形下的联合测量可以通过低维情形的简单推广实现。同时，这一协议还能满足类似的trade-off关系，并且对任意满足完美隐形传态的信道都适用？
%值得注意的是，本方案在推广至高维信道时存在局限性。其主要挑战源于随着系统维度$d_a$增加，正交子空间分解的数量呈指数增长（约∼$2^{d_a/2-1}-1$），导致Alice测量基的构建复杂度急剧上升。这引出了一个核心问题：是否存在一类隐形传态协议能够（i）通过低维情形的简单推广实现高维联合测量，（ii）同时保持类似的资源权衡关系，并且（iii）对所有支持完美隐形传态的信道都具有普适性？解决这一问题将成为未来研究的重要方向。
Notably, our scheme faces limitations when extended to higher-dimensional channels. The primary challenge stems from the exponentially growing complexity (scaling as $\sim2^{d_a/2-1}$) in constructing Alice's measurement basis, as the number of orthogonal subspace decompositions increases with the system dimension $d_a$. This raises a fundamental question: Does there exist a class of teleportation protocols where (i) high-dimensional joint measurements can be realized through simple low-dimensional extensions, while (ii) maintaining analogous resource trade-offs, and (iii) preserving universality for all perfect-teleportation-enabled channels? Addressing this question represents a promising direction for future research.

\section*{Acknowledgement}

J. L. C. was supported by the Innovation Program for Quantum Science and Technology (Grant No. 2024ZD0301000), and the National Natural Science Foundation of China (Grant No. 12275136).
F. L. Z. was supported by the National Natural Science Foundations of China (Grants Nos. 11675119). 
%J. L. C. was supported by the National Natural Science Foundations of China (Grants Nos. 12275136 and 12075001). 
D. Z. was supported by the Nankai Zhide Foundations.

\textbf{Declaration of generative AI and AI-assisted technologies in the manuscript preparation process.} 
During the preparation of this work, the author(s) used DeepSeek in order to check English expressions and translate certain sentences from Chinese into English. After using this tool/service, the author(s) reviewed and edited the content as needed and take(s) full responsibility for the content of the published article.

\bibliography{CQRPTbib}% Produces the bibliography via BibTeX.

\end{document}